\documentclass[12pt]{amsart}
\usepackage{geometry}                
\usepackage{graphicx}
\usepackage{amssymb}
\usepackage{epstopdf}
\usepackage{url}
\usepackage[toc,page]{appendix}
\usepackage{vmargin}
\setmarginsrb{25mm}{10mm}
             {25mm}{10mm}
             {10mm}{10mm}
             {10mm}{12mm}

\usepackage{color}

\title[Annotation of useR Data for UbiquitOUs Systems]{Challenges in Annotation of useR Data for \\UbiquitOUs Systems: \\\small{Results from the 1st ARDUOUS Workshop} \\\small{(ARDUOUS 2017)}}
\author[Yordanova et al.]{Kristina Yordanova$^{1,2}$, Adeline Paiement$^{3}$, Max Schr\"oder$^{1}$, Emma Tonkin$^{2}$, Przemyslaw Woznowski$^{2}$, Carl Magnus Olsson$^{4}$, Joseph Rafferty$^{5}$, Timo Sztyler$^{6}$}

\begin{document}
\maketitle

\begin{tabular}{p{0.4\textwidth} p{0.6\textwidth}l}
\begin{flushleft}
\includegraphics[width=0.4\textwidth]{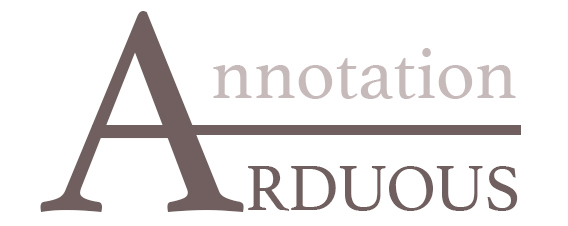}
\end{flushleft}
&
\begin{flushleft}
$^1$ University of Rostock, Rostock, Germany\\
$^2$ University of Bristol, Bristol, UK\\
$^3$ Swansea University, Swansea, UK\\
$^4$ Malm\"o University, Malm\"o, Sweden\\
$^5$ Ulster University, Ulster, UK\\
$^6$ University of Mannheim, Mannheim, Germany
\end{flushleft}\\
\end{tabular}

\section{Workshop Topic Description}
\vspace{0.1cm}

Labelling user data is a central part of the design and evaluation of pervasive systems that aim to support the user through situation-aware reasoning. It is essential both in designing and training the system to recognise and reason about the situation, either through the definition of a suitable situation model in knowledge-driven applications \cite{Yordanova:2015b,Chen:2014}, or through the preparation of training data for learning tasks in data-driven models \cite{Ordonez:2014}. Hence, the quality of annotations can have a significant impact on the performance of the derived systems.
Labelling is also vital for validating and quantifying the performance of applications. In particular, comparative evaluations require the production of benchmark datasets based on high-quality and consistent annotations.
With pervasive systems relying increasingly on large datasets for designing and testing models of users' activities, the process of data labelling is becoming a major concern for the community \cite{Alzahrani:2011}.


\textit{Labelling} is a manual process, often done by analysing a separately recorded log (video or diary) of the conducted trial. The resulting sequence of labels is called \textit{annotation}. It represents the underlying meaning of data \cite{Saldana:2016} and provides a symbolic representation of the sequence of events.

While social sciences have a long history in labelling (coding) human behaviour \cite{Saldana:2016}, coding in ubiquitous computing to provide the ground truth for collected data is a challenging and often unclear process \cite{Szewcyzk:2009}. It is usually the case that annotation processes are not described in detail and their quality is not evaluated, thus often making publicly available datasets and their provided annotations unusable \cite{Hein:2014}. Besides, most public datasets provide only textual labels without any semantic meaning (see \cite{Yordanova:2018a} for details on the types of annotation). Thus they are unsuitable for evaluating ubiquitous approaches that reason beyond the event's class and are able to provide semantic meaning to the observed data \cite{Yordanova:2015b}.

In addition, the process of labelling suffers from the common limitations of manual processes, in particular regarding reproducibility, annotators' subjectivity, labelling consistency, and annotator's performance when annotating longer sequences. Annotators typically require time-consuming training \cite{Hasan:2016}, which has as its goals teaching standardised ``best practice'' or increasing reliability and efficiency. Even so, disagreements between annotators, either semantically, temporally, or quantitatively, can be significant. This is often due to no real underlying ``ground truth'' actually existing, because of inherent ambiguities in annotating human activities, behaviours, and intents. Concurrent activities are also non-trivial to deal with and may be approached in multiple ways. The manual annotation approach is also unsuitable for in-the-wild long-term deployment, and methods need to be developed to help labelling in a big data context.

This workshop aims to address these problems by providing a ground for researchers to reflect on their experiences, problems and possible solutions associated with labelling. It covers 1) the role and impact of annotations in designing pervasive applications, 2) the process of labelling, the requirements and knowledge needed to produce high quality annotations, and 3) tools and automated methods for annotating user data.


To the best of our knowledge, no workshop or conference have yet focused on the complexity of the annotation process in its entirety, collating all of its aspects to examine its central role and impact on ubiquitous systems. We believe that this is indeed an important topic for the community of pervasive computing, which has been too often subsumed into discussions of related topics. We wish to raise awareness on the importance of, and challenges and requirements associated with high-quality and re-usable annotations. We also propose a concerted reflexion towards establishing a general road-map for labelling user data, 
which in turn will contribute to improving the quality of pervasive systems and the re-usability of user data. 

The remainder of this report will discuss the results of the ARDUOUS 2017 workshop.
The workshop itself had the following form: 
\begin{itemize}
\item a keynote talk addressing the problem that good pervasive computing studies require laborious data labeling efforts and sharing the experience in activity recognition and indoor positioning studies (see \cite{Maekawa:2017});
\item a short presentation session where all participants presented teasers of their work;
\item a poster session where the participants had the opportunity to present their work and exchange ideas in an informal manner;
\item a live annotation session where the participants used two annotation tools to label a short video; 
\item a discussion session where the participants discussed the challenges and possible solutions of annotating user data also in the context of big data.
\end{itemize}

Section \ref{sec:papers} will list the papers presented at the workshop.
The live annotation session of the workshop compared two annotation tools (Section \ref{sec:annotation}).
Part of the workshop was a discussion session, whose results are presented in Section \ref{sec:discussion}.
The report concludes with a short discussion of future perspectives in Section \ref{sec:conclusion}.
%
%

\section{Presented Papers}
\label{sec:papers}

Eight peer reviewed papers were presented at the workshop addressing the topics of annotation tools, methods and challenges. 
Additionally, one keynote on labelling efforts for ubiquitous computing applications was presented. 
Below is the list of presented papers and their abstracts.

\subsection{Good pervasive computing studies require laborious data labeling efforts: Our experience in activity recognition and indoor positioning studies \cite{Maekawa:2017}}

\paragraph{\textbf{Keynote speaker:}} Takuya Maekawa

\paragraph{\textbf{Abstract:}} \textit{Preparing and labeling sensing data are necessary when we develop state-of-the-art sensing devices or methods in our studies. 
Since developing and proposing new sensing devices or modalities are important in the pervasive computing and ubicomp research communities, we need to provide high quality labeled data by making use of our limited time whenever we develop a new sensing device. 
In this keynote talk, we first introduce our recent studies on activity recognition and indoor positioning based on machine learning.
Later, we discuss important aspects of producing labeled data and share our experiences gathered during our research activities.}

\subsection{A Smart Data Annotation Tool for Multi-Sensor Activity Recognition \cite{Diete:2017}}

\paragraph{\textbf{Authors:}} Alexander Diete, Timo Sztyler, Heiner Stuckenschmidt

\paragraph{\textbf{Abstract:}} \textit{Annotation of multimodal data sets is often a time consuming and a challenging task as many approaches require an accurate labeling. This includes in particular video recordings as often labeling exact to a frame is required. For that purpose, we created an annotation tool that enables to annotate data sets of video and inertial sensor data. However, in contrast to the most existing approaches, we focus on semi-supervised labeling support to infer labels for the whole dataset. More precisely, after labeling a small set of instances our system is able to provide labeling recommendations and in turn it makes learning of image features more feasible by speeding up the labeling time for single frames. We aim to rely on the inertial sensors of our wristband to support the labeling of video recordings. For that purpose, we apply template matching in context of dynamic time warping to identify time intervals of certain actions. To investigate the feasibility of our approach we focus on a real world scenario, i.e., we gathered a data set which describes an order picking scenario of a logistic company. In this context, we focus on the picking process as the selection of the correct items can be prone to errors. Preliminary results show that we are able to identify 69\% of the grabbing motion periods of time.}

\subsection{Personal context modelling and annotation \cite{Giunchiglia:2017}}

\paragraph{\textbf{Authors:}} Fausto Giunchiglia, Enrico Bignotti, Mattia Zeni

\paragraph{\textbf{Abstract:}} \textit{Context is a fundamental tool humans use for understanding their environment, and it must be modelled in a way that accounts for the complexity faced in the real world. Current context modelling approaches mostly focus on a priori defined environments, while the majority of human life is in open, and hence complex and unpredictable, environments. We propose a context model where the context is organized according to the different dimensions of the user environment. In addition, we propose the notions of endurants and perdurants as a way to describe how humans aggregate their context depending either on space or time, respectively. To ground our modelling approach in the reality of users, we collaborate with sociology experts in an internal university project aiming at understanding how behavioral patterns of university students in their everyday life affect their academic performance. Our contribution is a methodology for developing annotations general enough to account for human life in open domains and to be consistent with both sensor data and sociological approaches.}

\subsection{Talk, text or tag? The development of a self-annotation app for activity recognition in smart environments \cite{Woznowski:2017}}

\paragraph{\textbf{Authors:}} Przemyslaw Woznowski, Emma Tonkin, Pawel Laskowski, Niall Twomey, Kristina Yordanova, Alison Burrows

\paragraph{\textbf{Abstract:}} \textit{Pervasive computing and, specifically, the Internet of Things aspire to deliver smart services and effortless interactions for their users. Achieving this requires making sense of multiple streams of sensor data, which becomes particularly challenging when these concern people’s activities in the real world. In this paper we describe the exploration of different approaches that allow users to self-annotate their activities in near real- time, which in turn can be used as ground-truth to develop algorithms for automated and accurate activity recognition. We offer the lessons we learnt during each design iteration of a smart-phone app and detail how we arrived at our current approach to acquiring ground-truth data ‘in the wild’. In doing so, we uncovered tensions between researchers’ data annotation requirements and users’ interaction requirements, which need equal consideration if an acceptable self-annotation solution is to be achieved. We present an ongoing user study of a hybrid approach, which supports activity logging that is appropriate to different individuals and contexts.}

\subsection{On the Applicability of Clinical Observation Tools for Human Activity Annotation \cite{Krueger:2017}}

\paragraph{\textbf{Authors:}} 	Frank Kr\"uger, Christina Heine, Sebastian Bader, Albert Hein, Stefan Teipel, Thomas Kirste

\paragraph{\textbf{Abstract:}} \textit{The annotation of human activity is a crucial prerequisite for applying methods of supervised machine learning. It is typically either obtained by live annotation by the participant or by video log analysis afterwards. Both methods, however, suffer from disadvantages when applied in dementia related nursing homes. On the one hand, people suffering from dementia are not able to produce such annotation and on the other hand, video observation requires high technical effort. The research domain of quality of care addresses these issues by providing observation tools that allow the simultaneous live observation of up to eight participants – dementia care mapping (DCM). We developed an annotation scheme based on the popular clinical observation tool DCM to obtain annotation about challenging behaviours. In this paper, we report our experiences with this approach and discuss the applicability of clinical observation tools in the domain of automatic human activity assessment.}

\subsection{Evaluating the use of voice-enabled technologies for ground-truthing activity data \cite{Woznowski:2017a}}

\paragraph{\textbf{Authors:}} Przemyslaw Woznowski, Alison Burrows, Pawel Laskowski, Emma Tonkin, Ian Craddock

\paragraph{\textbf{Abstract:}} \textit{Reliably discerning human activity from sensor data is a nontrivial task in ubiquitous computing, which is central to enabling smart environments. Ground-truth acquisi- tion techniques for such environments can be broadly divided into observational and self-reporting approaches. In this paper we explore one self-reporting approach, using speech-enabled logging to generate ground-truth data. We report the results of a user study in which participants (N=12) used both a smart-watch and a smart-phone app to record their activities of daily living using primarily voice, then answered questionnaires comprising the System Usability Scale (SUS) as well as open ended questions about their experiences. Our findings indicate that even though user satisfaction with the voice-enabled activity logging apps was relatively high, this approach presented significant challenges regarding compliance, effectiveness, and privacy. We discuss the implications of these findings with a view to offering new insights and recommendations for designing systems for ground-truth acquisition ’in the wild’.}

\subsection{Labeling Subtle Conversational Interactions \cite{Edwards:2017}}

\paragraph{\textbf{Authors:}} Michael Edwards, Jingjing Deng, Xianghua Xie

\paragraph{\textbf{Abstract:}} \textit{The field of Human Action Recognition has expanded greatly in previous years, exploring actions and inter- actions between individuals via the use of appearance and depth based pose information. There are numerous datasets that display action classes composed of behaviors that are well defined by their key poses, such as ‘kicking’ and ‘punching’. The CONVERSE dataset presents conversational interaction classes that show little explicit relation to the poses and gestures they exhibit. Such a complex and subtle set of interactions is a novel challenge to the Human Action Recognition community, and one that will push the cutting edge of the field in both machine learning and the understanding of human actions. CONVERSE contains recordings of two person interactions from 7 conversational scenarios, represented as sequences of human skeletal poses captured by the Kinect depth sensor. In this study we discuss a method providing ground truth labelling for the set, and the complexity that comes with defining such annotation. The CONVERSE dataset it made available online.}			

\subsection{NFC based dataset annotation within a behavioral alerting platform \cite{Rafferty:2017}}

\paragraph{\textbf{Authors:}} Joseph Rafferty, Jonathan Synnott, Chris Nugent, Gareth Morrison, Elena Tamburini

\paragraph{\textbf{Abstract:}} \textit{Pervasive and ubiquitous computing increasingly relies on data-driven models learnt from large datasets. This learning process requires annotations in conjunction with datasets to prepare training data. Ambient Assistive Living (AAL) is one application of pervasive and ubiquitous computing that focuses on providing support for individuals. A subset of AAL solutions exist which model and recognize activities/behaviors to provide assistive services. This paper introduces an annotation mechanism for an AAL platform that can recognize, and provide alerts for, generic activities/behaviors. Previous annotation approaches have several limitations that make them unsuited for use in this platform. To address these deficiencies, an annotation solution relying on environmental NFC tags and smartphones has been devised. This paper details this annotation mechanism, its incorporation into the AAL platform and presents an evaluation focused on the efficacy of annotations produced. In this evaluation, the annotation mechanism was shown to offer reliable, low effort, secure and accurate annotations that are appropriate for learning user behaviors from datasets produced by this platform. Some weaknesses of this annotation approach were identified with solutions proposed within future work.}

\subsection{Engagement Issues in Self-Tracking: Lessons Learned from User Feedback of Three Major Self-Tracking Services \cite{Olsson:2017}}

\paragraph{\textbf{Authors:}} Carl M Olsson

\paragraph{\textbf{Abstract:}} \textit{This paper recognizes the relevance of self-tracking as a growing trend within the general public. As this develops further, pervasive computing has an opportunity to embrace user-feedback from this broader user group than the previously emphasized ‘quantified self:ers’. To this end, the paper takes an empirically driven approach to understand engagement issues by reviewing three popular self-tracking services. Using a postphenomenological lens for categorization of the feedback, this study contributes by illustrating how this lens may be used to identifying challenges that even best-case self-tracking services still struggle with.}

In summary, the workshop papers addressed both annotation tools and methodologies for providing high quality annotation as well as best practices and experiences gathered from self logging and annotation services and tools.  


\section{Live Annotation Session}
\label{sec:annotation}


The goal of the live annotation session was to identify challenges in the annotation of user activities. 
The identified challenges were used as a basis for the panel discussion entitled ``How to improve annotation techniques and tools to increase their efficiency and accuracy?".
The empirical data from the live annotation (the annotation itself and the questionnaire which can be found in the appendix) were used to empirically evaluate problems in annotating user data. 

%
%

\subsection{Annotation videos}

Originally, we have planned to annotate two types of behaviour: multi-user and single-user behaviour. 
Due to time constraints, at the end the participants annotated only the single user behaviour video\footnote{The time constraints were due to the fact that we underestimated the time needed for participants to become familiar with the tools. Additionally, the infrastructure requirements -- i.e. WiFi -- were a challenge to achieve reliably ``in the field'' annotation.}.

%
%
For the single user behaviour we used the CMU multimodal activity database\footnote{\url{http://kitchen.cs.cmu.edu/main.php}}.  
We selected a video showing the preparation of brownies. 
To show the complexity of annotating user actions, we followed two different action schemas: the first one being relatively simple and the second describing the behaviour on a more fine-grained level. 
The action schemas can be seen below.
\vspace{0.5cm}

\begin{tabular}{|l | p{12cm}|}
\hline
simple & clean, drink,  eat, get ingredients, get tools, move, prepare  \\
\hline
complex & clean, close, fill, open, put ingredients, put rest, put tools, shake, stir, take ingredients, take rest, take tools, shake, turn on, walk\\
\hline
\end{tabular}
%
%

\vspace{0.5cm}
The complex annotation schema contains some additional ``transition'' actions such as ``open'' to get something or ``turn on'' an appliance to enable another action. 

\subsection{Tools}

Two annotation tools were used during the annotation.

\subsubsection{MMIS Tool}
The MMIS annotation tool\footnote{ \url{https://ccbm.informatik.uni-rostock.de/annotation-tool/}} \cite{Schroeder:2016iwoar} is a browser-based tool and allows the annotation of single- and multi-user behaviour.
It allows both (1) online behaviour annotation and (2) annotation from video logs from multiple annotators in multiple tracks at the same time.
The tool is designed to be applicable in various domains and, thus, generic and flexible.
The annotation schema is read from a database, and possible constraints can be applied during the annotation.
This could ensure, for example, that the annotation is causally correct, or that there are no gaps between labels etc. (depending on the defined constraints). 
\begin{figure}[!h]
     \centering
    \includegraphics[width=1.0\textwidth]{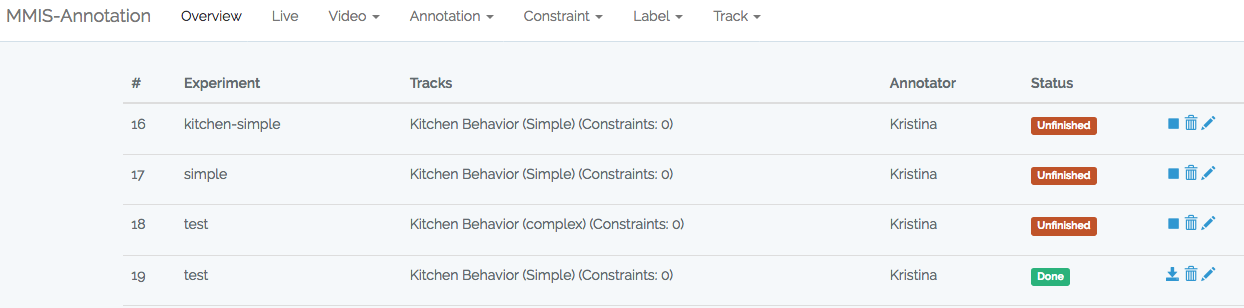} 
      \caption{The web interface of the MMIS tool.}
      \label{fig:toolMMIS}
\end{figure}

%
%
Figure \ref{fig:toolMMIS} shows the web interface of the MMIS annotation tool. 
The status of different annotation experiments can be seen, as well as the annotation constraints.  
More details about the tool can be found in \cite{Schroeder:2016iwoar}.

\subsubsection{SPHERE Tool}
The SPHERE annotation tool\footnote{\url{https://uob-sphere.onlinesurveys.ac.uk/arduous-workshop-downloads}} \cite{Woznowski:2017} is an android application and allows the annotation of single-user behaviour or self-annotation. 

Although the tool is not able to annotate multi-user behaviour, similarly to the MMIS tool it provides online annotation and uses rules to define the possible labels for a given location. 
In that manner, the tool provides location-based semantic constraints.  

\begin{figure}[!h]
     \centering
    \includegraphics[width=0.6\textwidth]{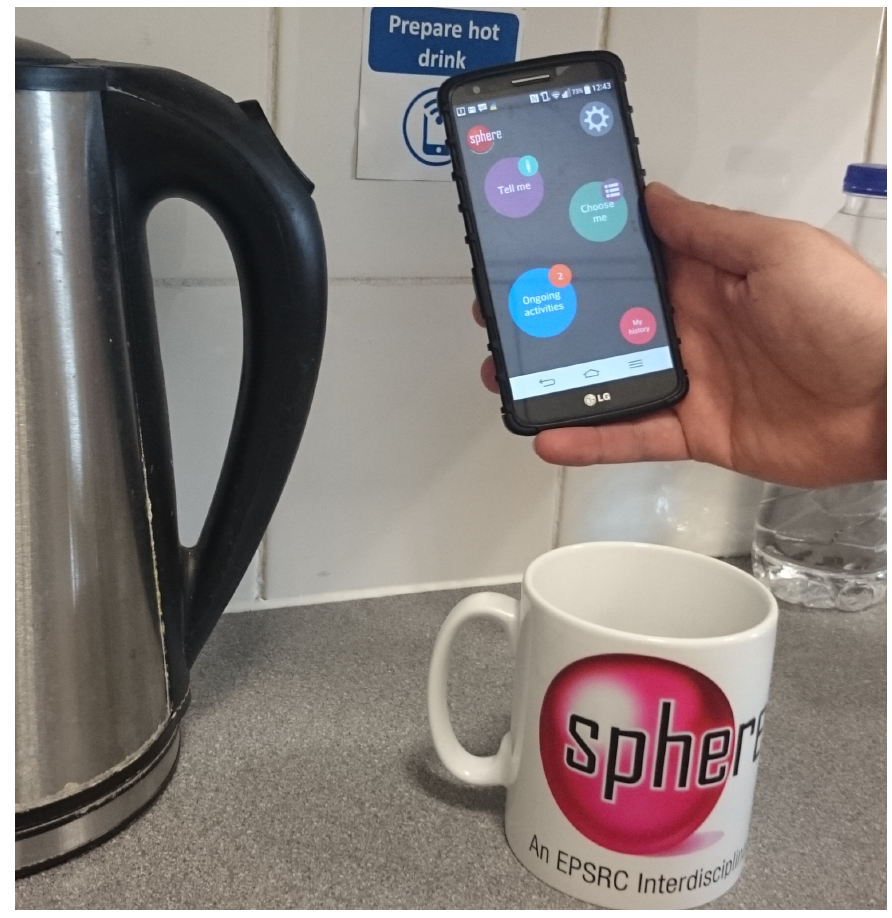} 
      \caption{The Android interface of the SPHERE tool.}
      \label{fig:toolSPHERE}
\end{figure}

Figure \ref{fig:toolSPHERE} shows the mobile interface of the SPHERE tool. 
The tool has an interactive interface, which allows selecting a given location where the action is happening and then selecting from a list of actions to annotate. 
More details about the tool can be found in \cite{Woznowski:2017}. 

\subsection{Participants}

\begin{figure}[!h]
    \includegraphics[width=1.0\textwidth]{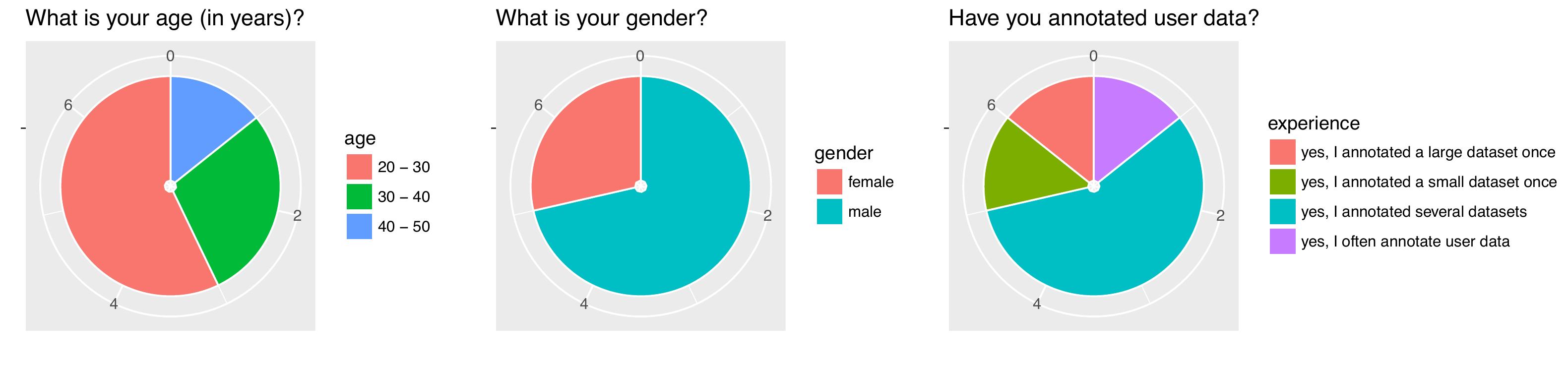}  
      \caption{Age, gender, and annotation expertise of participants.}
      \label{fig:age}
\end{figure}

Seven participants took part in the live annotation session. Figure \ref{fig:age}  (left and middle) shows the age and gender of the participants.
It can be seen that most of the participants were between 20 and 30 years old and that there were five male and 2 female participants.

All participants listed different fields of computer science, with many of them giving multiple fields of expertise. These were:
\begin{itemize}
\item active and assisted living;
\item activity recognition;
\item context-aware computing;
\item data analysis;
\item data bases;
\item human computer interaction; 
\item indoor positioning;
\item machine learning;
\item pervasive computing;
\item semantics;
\item ubiquitous computing.
\end{itemize}

%
%
Most of the listed areas can be considered as part of the ubiquitous computing topics.
When asked about their expertise in data annotation, 4 of them said they have annotated several datasets so far,  one often annotates datasets, one has annotated a small dataset once, and one has annotated a large dataset once (see Figure \ref{fig:age}, right). 
In other words, all participants had some experience with data annotation.

\subsection{Results}

\subsubsection{Questionnaire}
%
%
The questionnaire we used for the live annotation session can be found in Appendix A. 
Figure \ref{fig:anno} shows the results from the questionnaire regarding the choice of labels.
We have to point out that some of the participants only partially filled in the questionnaire, for that reason in some of the figures there are less than 7 samples.
It can be seen that according to participant feedback, the complex annotation schema labels were regarded as ``relatively clear'', while that of the simple schema was between relatively clear and clear (Figure \ref{fig:anno}, top left). 
Regarding the ease of recognising the labels in the video, one participant found it difficult to recognise them when using the simple schema, and 2 when using the complex schema. 
Three participants said that it was relatively easy to recognise the labels in the video (Figure \ref{fig:anno}, top right).
When asked whether the labels covered the observed actions, the participants pointed that the simple annotation schema covered some to most of the actions, while the complex schema covered most to all of the actions (Figure \ref{fig:anno}, bottom).
The results from Figure \ref{fig:anno} show that on the one hand the simple annotation schema was easier to understand and to discover in the video logs. 
\begin{figure}[!h]
      \begin{tabular}{c c}
    \includegraphics[width=0.47\textwidth]{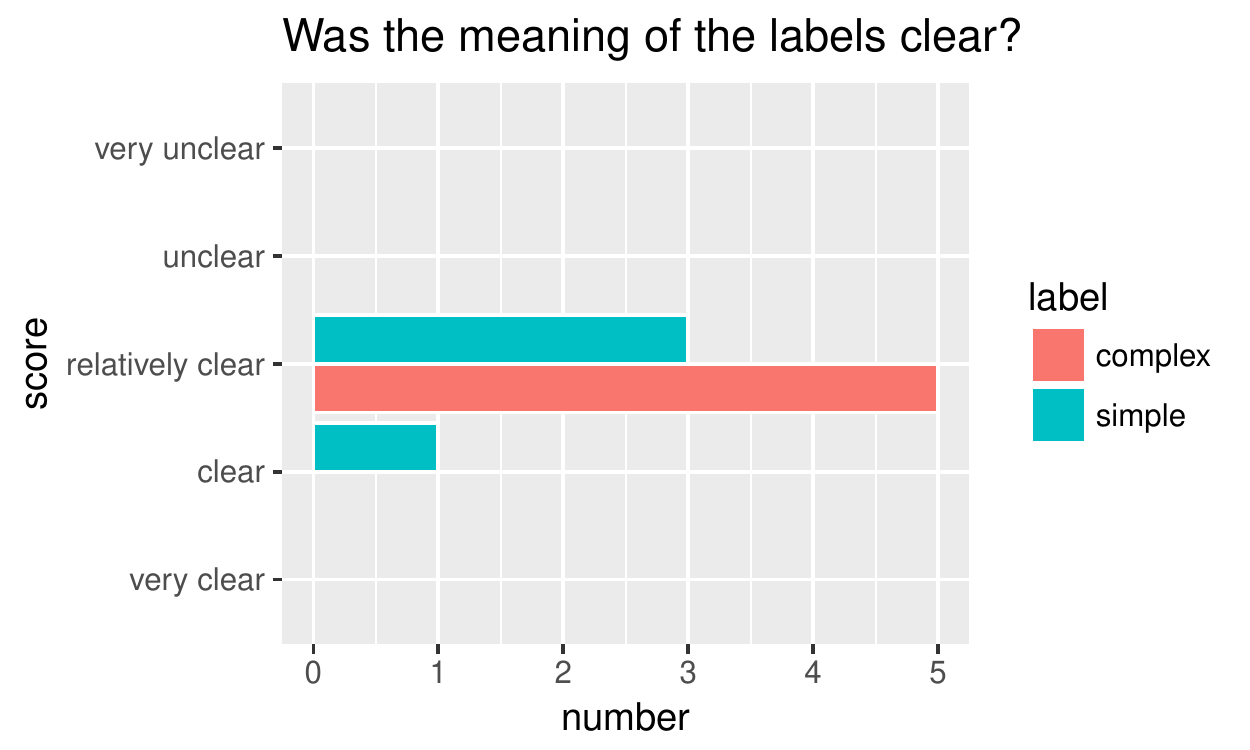} &  \includegraphics[width=0.53\textwidth]{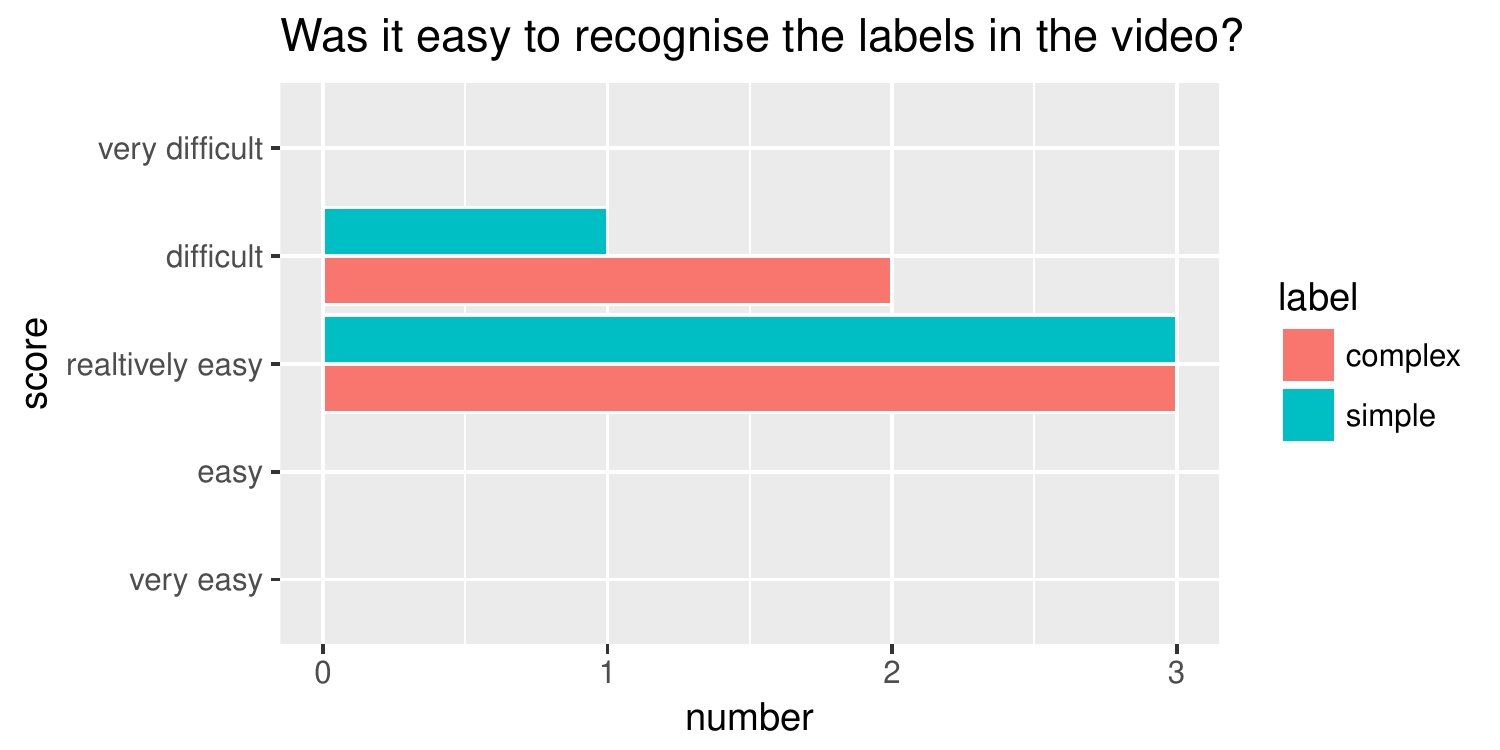}\\
    \multicolumn{2}{c}{ \includegraphics[width=0.53\textwidth]{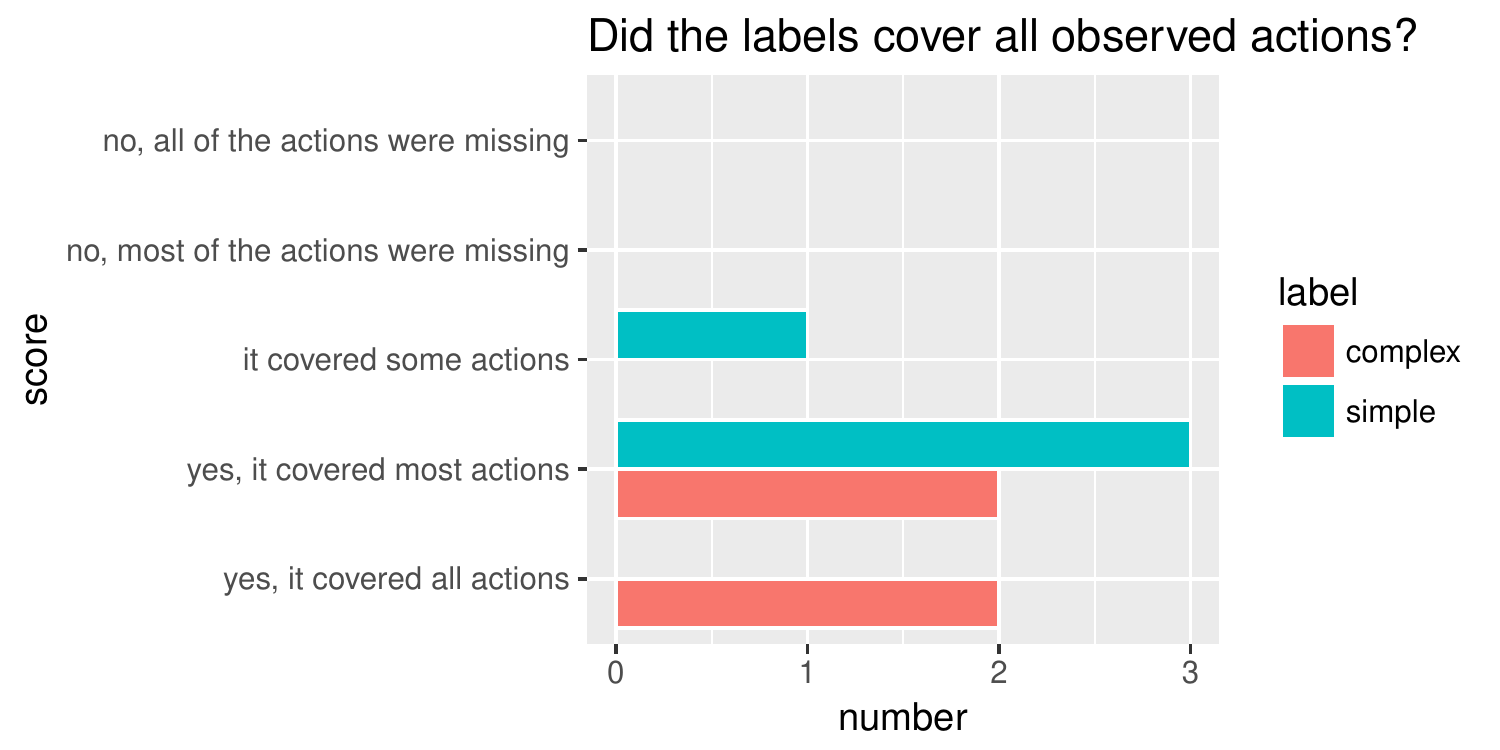}}\\
    \end{tabular}
      \caption{Evaluation of selected labels.}
      \label{fig:anno}
\end{figure}
On the other hand, the complex annotation schema covers more fully the actions observed in the video logs. 
This indicates that there is a trade-off between simplicity and completeness. 
Depending on the application, one should find the middle ground between the two. 

\begin{figure}[!h]
     \centering
    \includegraphics[width=1.0\textwidth]{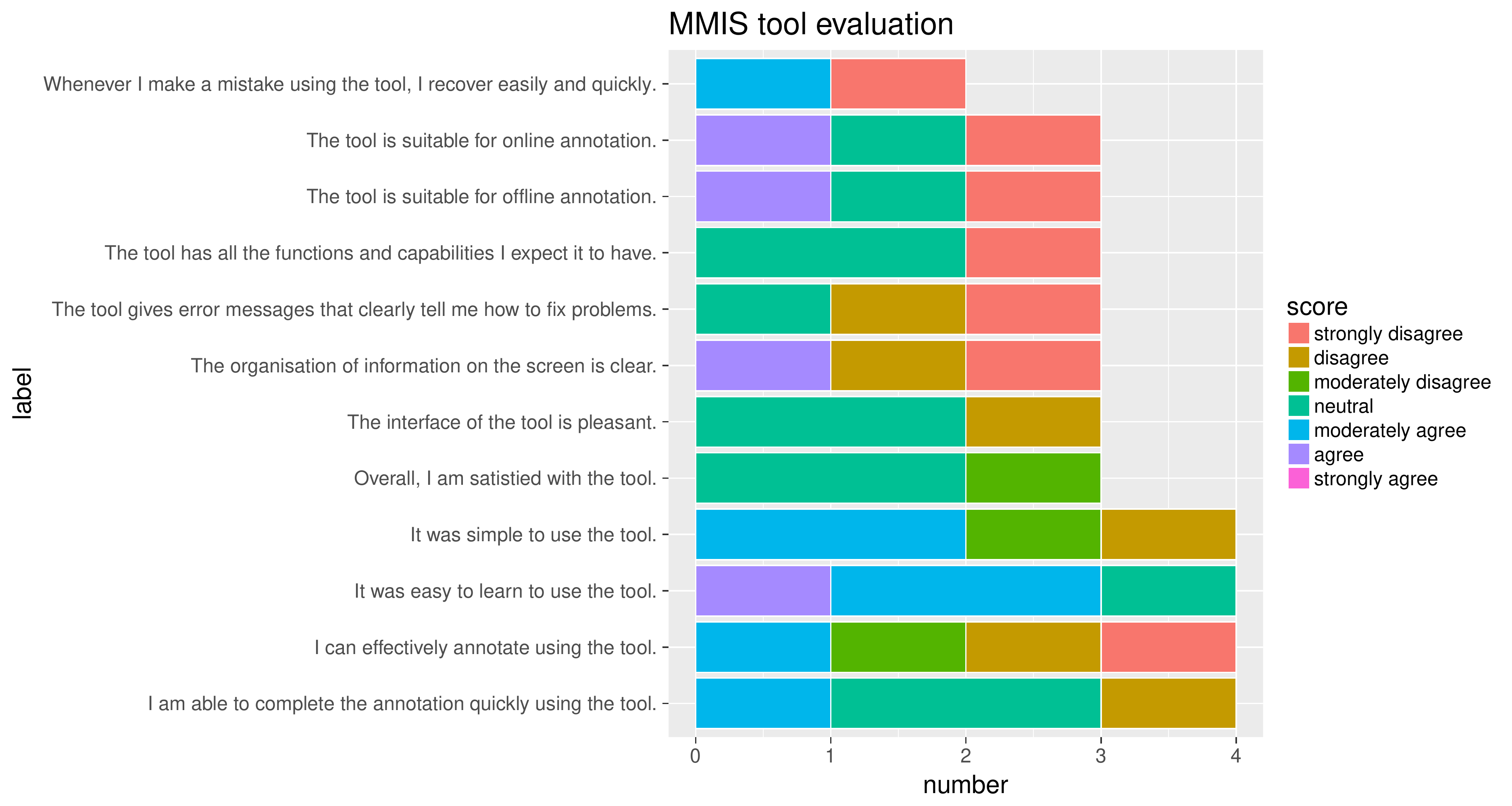} 
      \caption{Evaluation of the MMIS annotation tool.}
      \label{fig:annoMMIS}
\end{figure}

\begin{figure}[!h]
     \centering
    \includegraphics[width=1.0\textwidth]{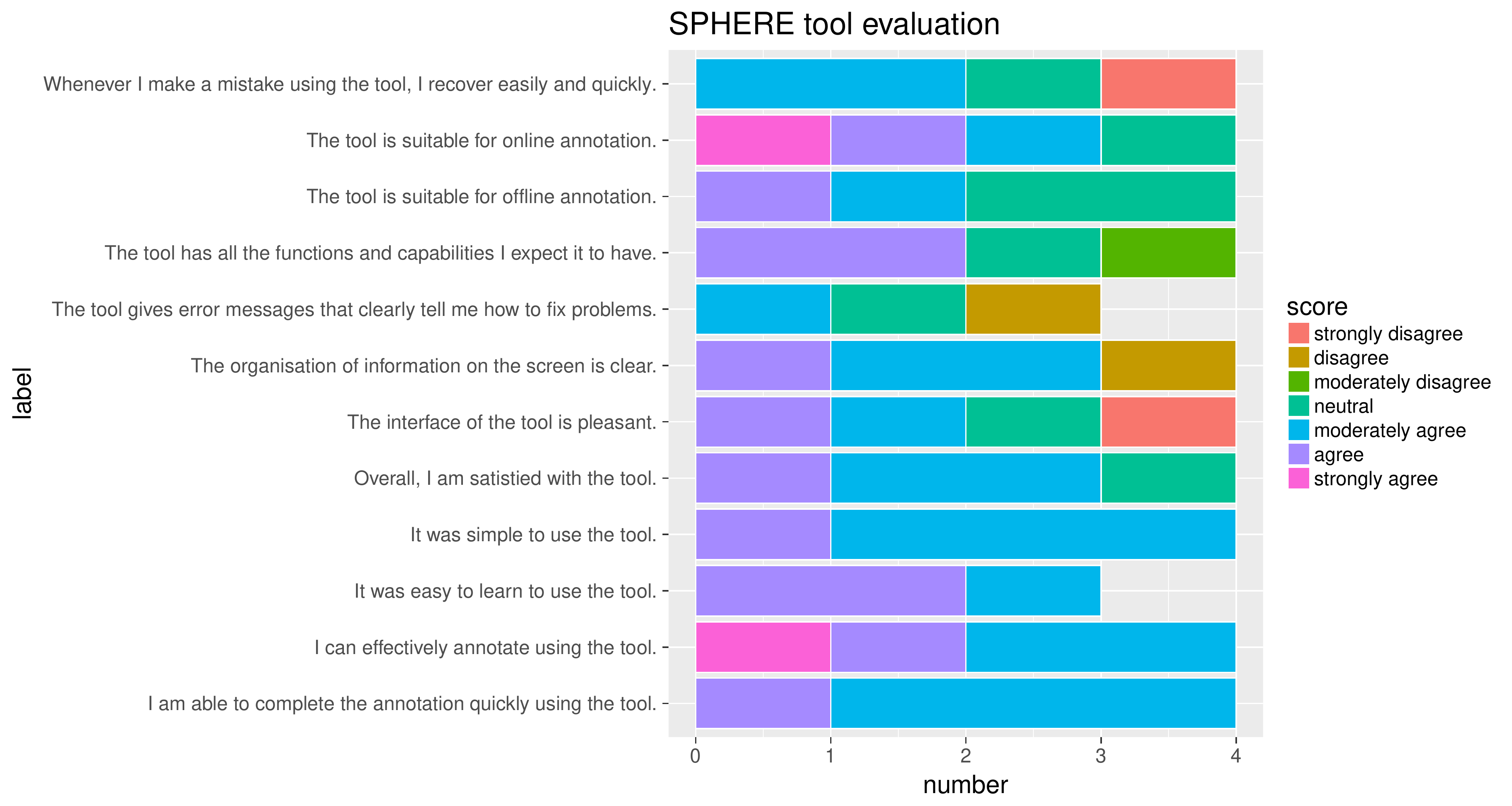} 
      \caption{Evaluation of the SPHERE annotation tool.}
      \label{fig:annoSPHERE}
\end{figure}

Figure \ref{fig:annoMMIS} and Figure \ref{fig:annoSPHERE} show the results from the questionnaire addressing the two annotation tools. 
Four of the participants used the SPHERE tool and three used the MMIS tool. 
One of the participants started using the MMIS tool, then switched to the SPHERE tool\footnote{No reason was given for switching the tools.}, for that reason there are some answers with sample of 4  in Figure \ref{fig:annoMMIS}.
%
%
As with the labels evaluation, here too some of the participants did not answer all questions.
For that reason, at places we have a sample $< 4$, $< 3$ respectively. 
From both figures, it can be seen that the SPHERE tool received better scores than the MMIS tool. 
Overall, the tool was easier to use, it was better suited for online annotation, and the participants were able to effectively annotate with the tool.
This could potentially indicate that tools designed for a specific domain are better suited for labelling than generic tools such as the MMIS tool.
%
%

Beside the Likert scale questions, some free text questions were asked which are analysed below. An example questionnaire with all questions can be found in Appendix A.

\paragraph{\textbf{Annotation problems.}} Among the annotation problems the participants experienced were ambiguous labels, annotation of simultaneous actions, the quality of the video and its speed (the video was too fast and unstable to notice the executed action\footnote{What is meant here is that we used the video from the head mounted camera for the annotation. This means the camera was not fixed at a certain position and when the person moved, the video looked unstable.}), inability to see when the action starts and ends, difficulty in remembering all available action labels (especially in the SPHERE tool, one had to scroll to see them).  
%
%

\paragraph{\textbf{Missing tool functionality.}} The MMIS tool did not have option for replaying the annotation sequence in order to correct the labels and time details were missing. 
The SPHERE tool also did not provide timelines for offline annotation. 
One participant pointed out that a statistical suggestion of the next label to annotate would have been desirable.  

\paragraph{\textbf{Tool advantages.}} The MMIS tool had the advantages of having a straight-forward implementation, online option, as well as an option to load data based on which to annotate. 
The SPHERE tool had the advantages of working basic functionality and ease of use.  

\paragraph{\textbf{Tool disadvantages.}} Although the MMIS tool had a lot of desirable features, according to a participant the interaction design needed a complete change. 
The SPHERE tool had the disadvantage of difficulty to see the vocabulary as one had to scroll to see all labels. 

\subsubsection{Annotation}



\begin{figure}[!h]
     \centering
    \includegraphics[width=1.0\textwidth]{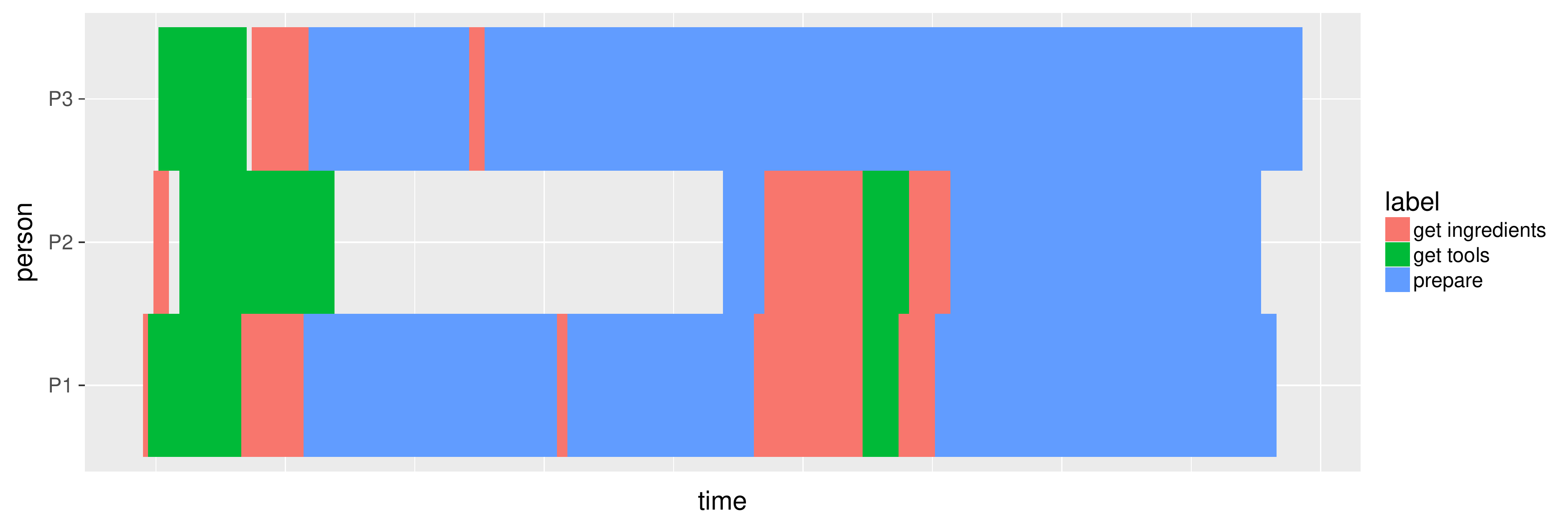} 
      \caption{Annotation during the live annotation session with the MMIS annotation tool and the simple annotation schema.}
      \label{fig:annoMMIS-simple}
\end{figure}

\begin{figure}[!h]
     \centering
    \includegraphics[width=1.0\textwidth]{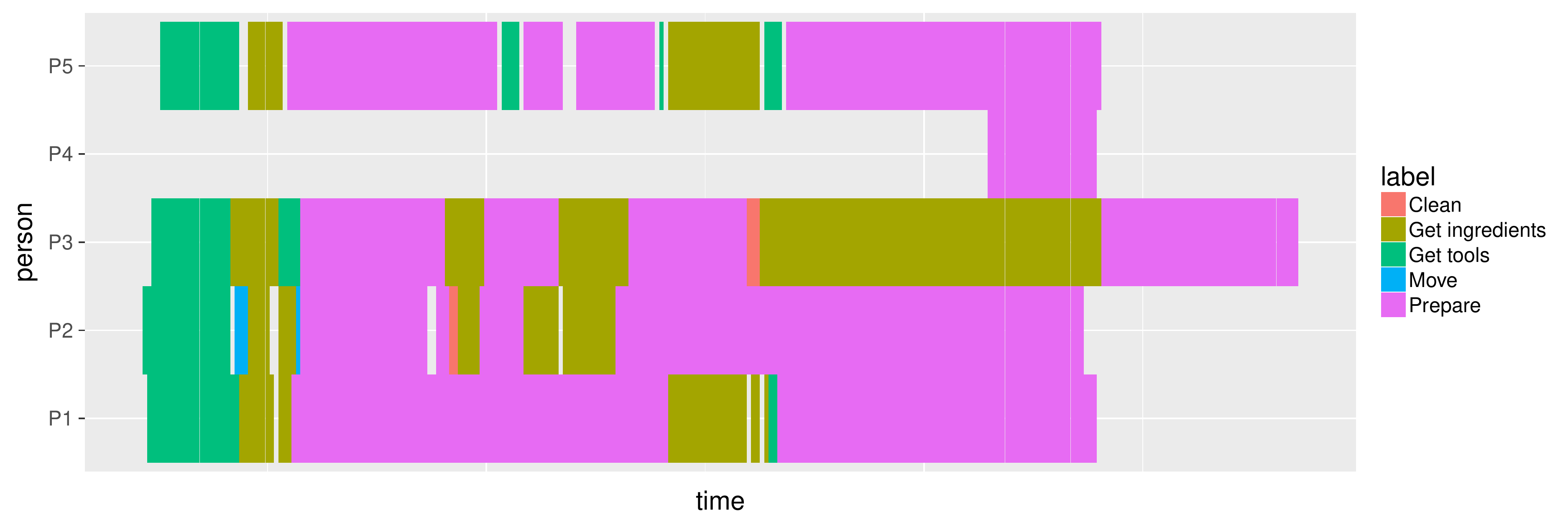} 
      \caption{Annotation during the live annotation session with the SPHERE annotation tool and the simple annotation schema.}
      \label{fig:annoSPHERE-simple}
\end{figure}

During the annotation session, 3 participants annotated the excerpt of the CMU video log with the MMIS tool. 
5 participants annotated the video log with the SPHERE tool. 
In other words, one participant did not fill in the questionnaire after annotating with the SPHERE tool. 
Figure \ref{fig:annoMMIS-simple} shows the annotation performed with the MMIS tool and the simple annotation schema, while Figure \ref{fig:annoSPHERE-simple} shows the same annotation schema with the SPHERE tool. 
It can be seen that the participants using the MMIS tool used three labels for the annotation (\textit{get ingredients}, \textit{get tools}, and \textit{prepare}). 
In the annotation produced with the SPHERE tool, one participant additionally used the label \textit{move} and two participants used the label \textit{clean}.

%
%
When looking at the overlapping, both figures show that there were serious discrepancies between the labels of the different annotators. 
Some of the observed problems are gaps in the annotation, inability to detect start and end of action, and different interpretation of the observed actions. 
Apart from the gaps, these problems were also mentioned in the questionnaire.  

\begin{figure}[!h]
     \centering
    \includegraphics[width=1.0\textwidth]{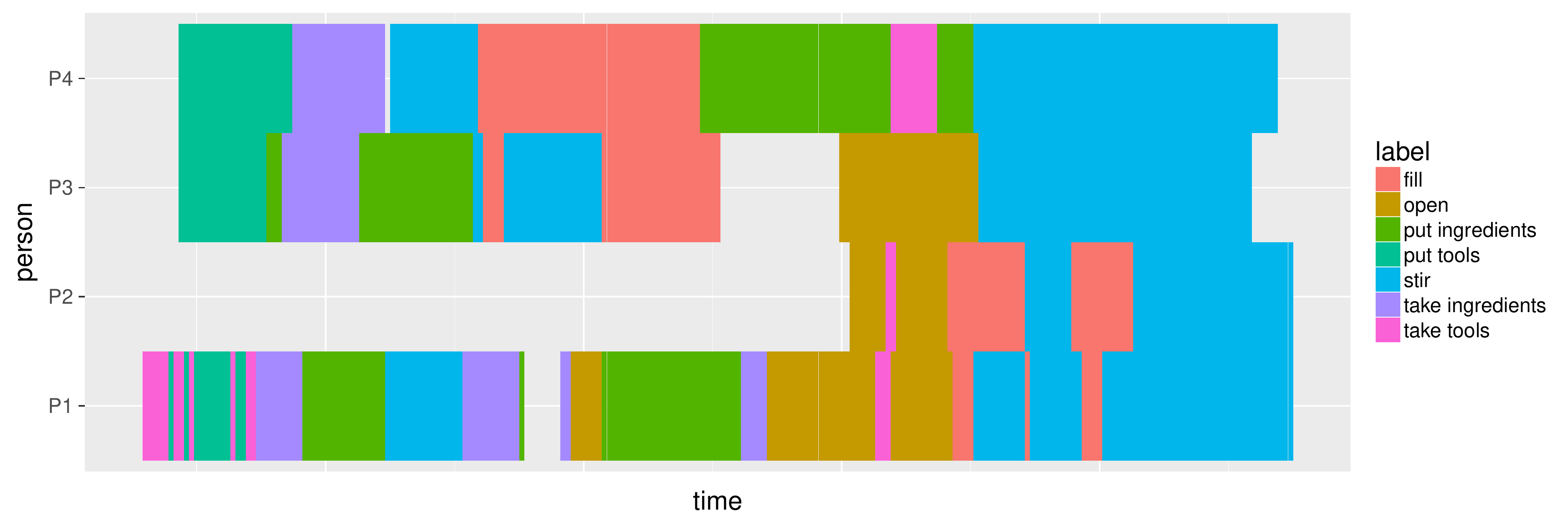} 
      \caption{Annotation during the live annotation session with the MMIS annotation tool and the complex annotation schema.}
      \label{fig:annoMMIS-complex}
\end{figure}

\begin{figure}[!h]
     \centering
    \includegraphics[width=1.0\textwidth]{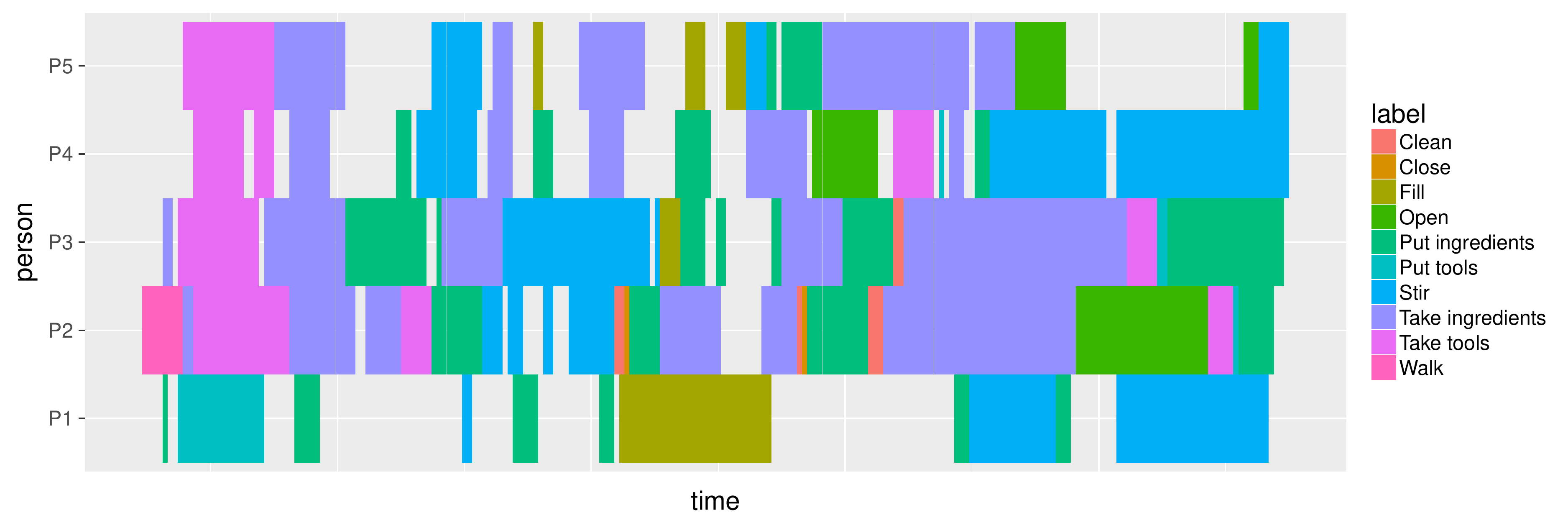} 
      \caption{Annotation during the live annotation session with the SPHERE annotation tool and the complex annotation schema.}
      \label{fig:annoSPHERE-complex}
\end{figure}

Similar problems were also observed in the annotation with the complex activities. 
Figure \ref{fig:annoMMIS-complex} shows the annotation with the MMIS tool and the complex annotation schema, while Figure \ref{fig:annoSPHERE-complex} shows the same annotation schema with the SPHERE tool.
Here too, the participants using the MMIS tool tended to use less labels than the participants using the SPHERE tool (7 labels in the case of the MMIS tool and 10 in the case of the SPHERE tool).
Furthermore, with both annotation schema, the users of the SPHERE tool tended to produce more actions than those using the MMIS tool. 
This also had effect on the length of the actions, with the users of the SPHERE tool producing shorter actions than the users of the MMIS tool.  
One potential explanation for this could be that the SPHERE tool allows easier switching between activities. 
Another explanation, however, could be related to the fact that in the SPHERE tool one has to scroll to see all the labels, which could lead to assigning incorrect labels and then quickly switching to the correct one. 

The above observations indicate that future live annotation tools should be aware of and try to cope with problems such as gaps in the annotation, inability to detect start and end of action, and different interpretation of the observed actions.

One important aspect, one should take into account, is that the participants did not have any previous training on using the tool or recognising the labels.
Usually a training phase is performed, in which the annotators learn to assign the same labels for a given observed activity. 
The learning phase is also used for other aspects such as to learn to recognise start and end of a given activity. 
%
%
When annotating the CMU dataset, recent results have shown that two annotators require to annotate and discuss about 6 to 10 videos before they are able to reach a converging interrater reliability \cite{Yordanova:2018a}.  


We also looked into the interrater reliability between each two annotators for both tools\footnote{Note that the small sample size is not sufficient to generalise the findings. We however use it to illustrate the observed problems with this small sample size.}. 
We used Cohen's kappa to measure the reliability as it is an established measure for such kind of problems. 
A Cohen's kappa between 0.41 -- 0.60 means moderate agreement, between 0.61 -- 0.80 means substantial agreement, and above 0.81 indicates almost perfect agreement \cite{Landis:1977}.
\begin{figure}[!h]
     \centering
    \includegraphics[width=0.6\textwidth]{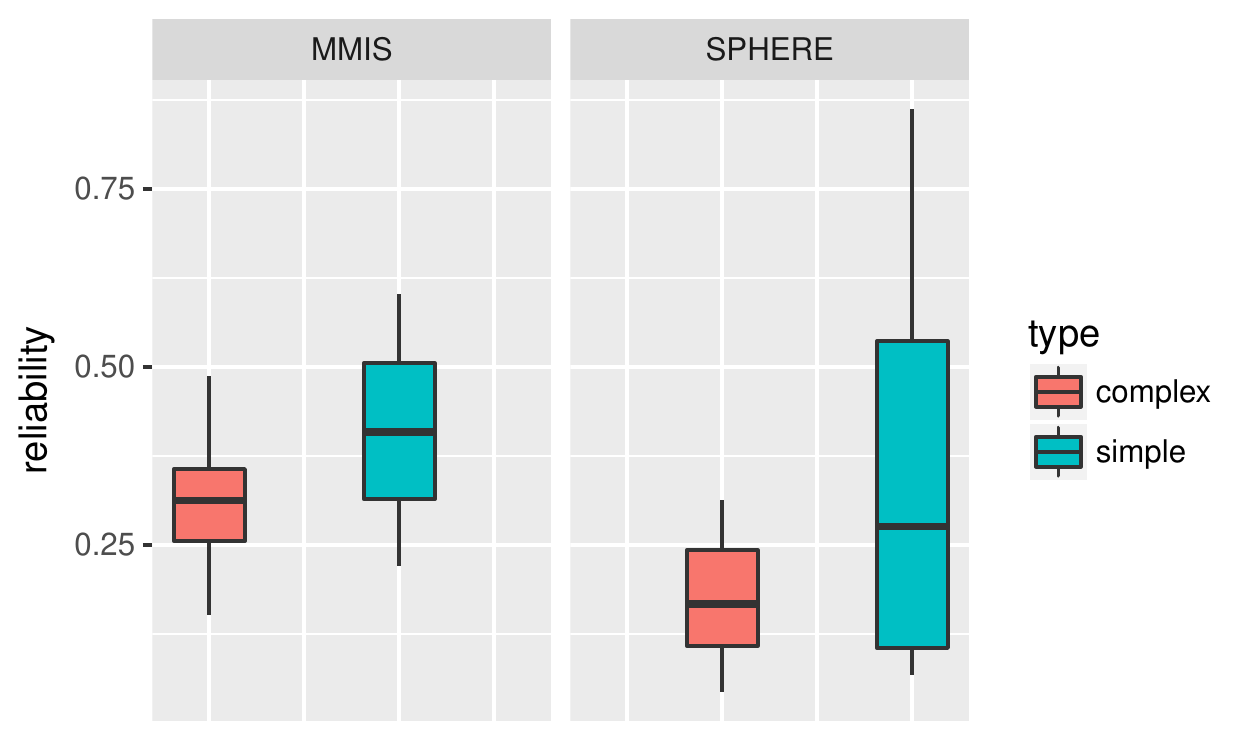} 
      \caption{Comparison between the median interrater reliability with the MMIS and SPHERE tools. Cohen's kappa was used as reliability measure.}
      \label{fig:comparison-reliability}
\end{figure}
Figure \ref{fig:comparison-reliability} shows the comparison between the median interrater reliability for the simple and complex annotation schema for both tools. 
It can be seen that the annotation with the complex schema produced lower interrater reliability with a mean of $0.31$ for the MMIS tool and a mean of $0.17$ for the SPHERE tool. 
In comparison, the annotation with the simple annotation schema produced better results with a mean of $0.41$ for the MMIS tool and a mean of $0.34$ for the SPHERE tool.
Surprisingly, despite the worse usability evaluation of the MMIS tool, the agreement between the annotators was higher than with the SPHERE tool. 
This can, of course, be explained with the fact that the annotators using the MMIS tool used fewer labels than those using the SPHERE tool. 
%
%
Another problem is that the annotation with the SPHERE tool contains more gaps than the annotation with the MMIS tool. 
This could be due to the annotation but also due to the mechanism the SPHERE tool uses to record the labels\footnote{The MMIS tool had the constraint to avoid gaps between annotation, while the SPHERE tool did not have this constraint.}.


\section{Discussion Session}
\label{sec:discussion}

\subsection{Preliminaries}

Before presenting the results from the discussion session, below are some preliminary challenges and scenarios we used as basis for the discussion (beside the outcome of the live annotation session). 

\subsubsection{Known challenges in annotation of user data}
As discussed in the introduction, there are different known problems and challenges associated with annotation of user data. 
Some of them are listed below.
%
%
\begin{itemize}
\item	\textbf{Types of annotation schema:} it is not always clear and easy to determine what annotation schema should be used. One has to decide on the labels, their meaning, the information included in each label etc. For example, do we annotate the action as ``cook'' or ``stir''? Do we add the objects on which the action is executed? Do we include the particular location? 
\item	\textbf{Annotation granularity:} it is sometimes difficult to determine the annotation granularity. Is it defined by the sensors or by the problem domain \cite{sensorsGranularity:Chen}? Can we reason on the granularity level of annotation that is not captured by the sensors \cite{Yordanova:2015b}? For example, do we annotate the action ``prepare meal'' or the subactions ``take the ingredients'', ``turn on the oven'', ``boil water'' etc., especially if the sensors are able only to detect that the oven is on, but not if an ingredient is taken?  
\item	\textbf{The meaning behind the label:} the label provides only a class assignment but is it possible to reason about the semantic meaning behind this label \cite{Yordanova:2018a}? For example, the action ``cut the onion'' is just a string but for us as humans it has a deeper meaning: to cut we probably need a knife; the onion has to be peeled; the onion is a vegetable, etc.
\item	\textbf{Identifying the beginning and end of the annotation label:} the start and end of a given event can be interpreted in different ways, which changes the duration of the event and the associated sensor data. For example, in the action ``take the cup from the table'', does ``take'' start the moment the hand grasps the cup or the moment the cup is lifted off the table, or when the cup is already in the person's hand? 
\item	\textbf{Keeping track of multi-user behaviour:} especially in live annotation scenarios it is not easy to keep track of multiple users. Increasing the number of users makes it more and more difficult to follow the behaviour. For example, in assessment systems such as Dementia Care Mapping (DCM), the observer is able to annotate the behaviour of up to 8 persons simultaneously \cite{Kitwood:1992} but the interrater reliability is relatively low \cite{Krueger:2017}.
\item	\textbf{Keeping track of the identity of the users in multi-user scenario:} it is even more difficult to consistently keep track of the identity of each person when annotating multi-user data. Here we have the same problems as with the above point. 
\item	\textbf{Online vs. offline tools:} online tools allow annotation on the run but they are often not very exact as it is difficult to track the user's behaviour in real time. This was supported by the results from our live annotation session, where the interrater reliability was very low especially when using more complex annotation schemas. On the other hand, offline tools can produce high quality annotation with almost perfect overlapping between the annotators (e.g. \cite{Yordanova:2018a}) but it costs a considerable amount of time and effort and requires training the annotators. 
\item	\textbf{Time needed to annotate vs. precision:} the precision is improved with the time spent for annotating but it is a tedious process  \cite{Yordanova:2018a} and good annotators are difficult to find / train. 
\end{itemize}

\subsubsection{Running scenario}
As start of the discussion session, we considered the following scenario. It describes a hypothetical effort of collecting and annotating data that describes the vocabulary and semantic structure of health dynamics across different individuals and countries.  

\textit{How about if you started with acquiring real life action information of 1000 persons over 1 year in 1 country, annotated it with ground truth information on the activities, and come up with the action vocabulary and grammar for that country. This may then be used to determine the healthy aging-supportiveness of a given country and the possibility space of a given individual. Then you repeat this for 100 more countries.}

From this scenario the following questions arise:
\begin{itemize}
\item What would be needed to complete such annotation within 3-4 years? 
\item How could the results of such an investment then be used?
\item	What are the challenges at obtaining the ground truth?
\item	What is the tradeoff between efficiency and quality?
\item	Is it possible at all to annotate such amount of data manually?
\item	Could smart annotation tools and machine learning methods help to annotate large amounts of user data?
\item	What is the role of the user interaction for improving the performance of the annotation (especially in the case of automatic annotation)?
\end{itemize}

\subsection{Results from the discussion}
Based on the questions identified in the preliminaries, the problems discovered in the live annotation session, and the participants' experience in annotation, the following challenges and possible solutions were identified. 

\subsubsection{Improving our annotation tools}

Even though several annotation tools are available online \cite{elan:2009,Do:2016,Hollen:2014,Vondrick:2012,Hagedorn:2008}, they have a number of limitations, and do not present some key desirable functionality, especially towards improving quality.
The following desirable functionality to improve annotation quality has been identified: 
\begin{itemize}
\item \textbf{automatic control of embedded constraints}, such as no gaps between labels, causal correctness, etc.;
%
%
\item the ability of the user to \textbf{add missing labels} (i.e. the tool allows editing past annotation);
%
%
\item going back to the annotation and \textbf{editing} it (caution: the suitability of this feature depends on the type and motivation of the annotator);
\item in order to improve the annotation accuracy, \textbf{display all available data channels} at the same time, not just the video;
\item verification of the performance should not use the same data. A \textbf{leave-some-out strategy} would be acceptable;
\item when it is obvious that the annotation of a sequence will be of \textbf{low quality}, for example due to the tool freezing, the \textbf{annotator could signal} this;
\item \textbf{grouping tags for mental mapping} would allow a faster and more intuitive use of the tool;
\item some level of \textbf{automation} may come from using \textbf{special sounds} to identify actions, such as the toilet flushes sound.
\end{itemize}

Some special considerations for an online annotation tool, necessary to keep up with fast pace live annotation, are listed below: 
\begin{itemize}
\item latency management; 
\item speed optimisation.
\end{itemize}
Improving annotation tools is one approach to improving the annotation quality: better documentation, training, interface, etc. But equally important might be a bit of machine learning, however unreliable, which could possibly help out with improving the annotation quality. For example, one could recommend annotations that are likely to follow previously set annotations on the basis of existing data. This could also potentially encourage annotators to consider likely/relevant annotations. 


\subsubsection{Design of ontologies}
Ontologies are used to represent the annotation schema and the semantic relations / meaning behind the used labels.
They are important for both building model driven systems and for evaluating data-driven approaches as the provide not only the label but also semantic information about its meaning and its relations to other labels as well as to unobserved phenomena. 
Below are some considerations regarding the design and use of ontologies for annotation of user data. 
\begin{itemize}
\item Building a model and \textbf{designing the related ontology should be done jointly} (involving the system designers, annotators and potential users) in order to obtain a suitable ontology;
\item discussing with, and accounting for the \textbf{feedback of the annotators} could help in improving the quality/suitability of the ontology (caution: here we potentially can run into the risk of over-fitting the annotation);
\item \textbf{subjectivity of labels:} many labels are subjective and will be interpreted in a different manner by different annotators (here a solution is using a \textbf{training phase} where the annotators learn to assign the same labels to the same observations. Another solution might be (even if this requires high effort) to make a \textbf{sample database} with video examples for each annotation or to add a detailed explanation about the meaning);
%
%
\item defining an ontology is not only defining a set of \textbf{labels} but also their \textbf{meaning}. Therefore it is necessary to go through several iterations. Another option could be to rely on existing \textbf{language taxonomies} \cite{Miller:1995} to define the underlying semantic structure; 
\item \textbf{group/consensus annotations:} we could let different annotators generate their own list of labels. We would then obtain a richer set of annotations. These annotations can then be grouped into a \textbf{hierarchical structure} where \textbf{synonymous labels} belong to the same group (note: looking at the different interpretations of people could potentially help narrowing down the definition of a label.). This however has the potential danger of losing specific meanings due to generalisation. 
\item an ontology is \textbf{not static}, it \textbf{evolves} with time (similar to how our behaviour changes with time). One should be aware that the ontology has to be updated and maintained over time. It is an open question of how often the ontology has to be updated.   
\end{itemize} 
As a side note, it would be interesting to see if the \textbf{interpretation of labels changes with time} together with the natural change in perception.

The ontology should enable the \textbf{purpose of a study}, but should also define actions that are identifiable in the available data. 
Therefore, its design should seek for a compromise solution between:
\begin{itemize}
\item what can we have? (data-based view);
\item what is useful? (purpose-based view).
\end{itemize}
This consideration is especially relevant when choosing the level of granularity.

\subsubsection{Large scale data collection}
Based on the running scenario for large scale data collection, the following aspects were discussed.
\begin{itemize}
\item In order to look for correlations between different individuals with various social and cultural background, two levels of labels are needed: \textbf{fine-grained activities} of daily living, and overall \textbf{ageing and health data};
\item the fine-grained activity recognition involves a huge amount of data to be collected, which poses the very challenging problems of \textbf{data transmission} and \textbf{annotation}, and of \textbf{privacy} and \textbf{security};
\item for such large scale data collection, the \textbf{purpose} of the collected data should also be well defined beforehand.
\end{itemize}

\textbf{Self-annotation.}
Obtaining annotations for big datasets is a particularly daunting task. 
This problem could potentially be solved by using self-annotation tools and techniques where the study participants are those who annotate their own data \cite{Woznowski:2017,Schroeder:2016iwoar}.

During the discussion, the self-annotation was seen as the only practical way of obtaining fine-grained activity labels -- although it is not a perfect solution on its own, as will be discussed later. 
Its use is facilitated by the already existing and popular life-logging apps \cite{Olsson:2017}.
The practice of self-annotation already exists in various commercial life-logging apps as well as research experiments \cite{annotation1,Woznowski:2017,Schroeder:2016iwoar,Olsson:2017}. 
Gaining access to these annotations would make the annotation gathering process much easier.

For people who do not normally log their actions, we need to come up with motivation strategies. 
The consensus is that subjects will be more willing to participate if they can benefit from the study. 
From experience, to let people \textbf{use the experiment equipment for personal purposes}, and eventually even to \textbf{allow them to keep it} after the experiment, seems to be an effective incentive. 
\textbf{Monetary compensation} is a good incentive as well, especially if combined with the ability to select the data that will be shared\footnote{For example, Google has a tool called Google Opinion Rewards where one could participate in surveys and in the process earn Google Play credit (see \url{https://play.google.com/store/apps/details?id=com.google.android.apps.paidtasks&hl=en}).}. 
\textbf{Gamification} was also tested by some participants and noted that it could potentially be used as motivation for annotating and sharing their data.

It is to be noted that some populations are easier to be recruited for self-annotation such as young, female, single, extreme cases (as opposed to elderly, married, exhibiting multiple health conditions). 
It may therefore be \textbf{difficult to obtain a representative sample of population} for the experiments if the participants are restricted to the ones willing (or able) to self-annotate. 
Similarly, the incentive and type of incentive may also induce a \textbf{selection bias}.

Another problem is that people tend to \textbf{change their behaviours} when / because they are monitored. 
The data collection may therefore be biased by this effect too, especially if the subject is self-annotating and self-aware.

In order to maintain a level of quality, self-annotation \textbf{should not be done in isolation}, and people who annotate should be \textbf{able to refer to an expert} to verify that the annotation is being done correctly.

In addition, an \textbf{expert should refine} the user's self-annotation -- without too large burden for the expert. 
This may be done \textbf{manually}, or be assisted by \textbf{comparing data} from other users and identifying outliers. 
It is also possible to ask annotators to \textbf{evaluate the labels produced by peers} -- with some privacy issues. 
For example, an expert could identify inputs that can be classified very reliably and which is a good proxy to an action/activity that is in itself not observed.
Missing or mis-placed labels associated to easily recognisable actions (e.g. flushing the toilets and its characteristic sound) could also be detected \textbf{automatically by a smart tool}\footnote{For example, Nokia uses automatic activity recognition, but offers users to re-label these activities if needed, as well as asks users whenever the activity is not recognised. This feedback is then used in the machine learning elements of the products to reduce how much self-annotation is needed (for more information see \url{https://support.health.nokia.com/hc/en-us/articles/115000426547-Health-Mate-Android-App-Activity-recognition}).}. 

The evaluation of self-annotation quality should be done well before the end of an experiment in order to provide \textbf{feedback} or \textbf{retraining} to the annotators if necessary.

Participants should be given the chance to provide \textbf{qualitative feedback} that can help improve the tools and ontologies. 
This would also help keeping them engaged.

\textbf{Purpose.}
During the discussion, it was identified that the \textbf{research question} plays an important role in the way the user data is collected and annotated.
Collecting data without research question was generally seen as contra productive: if one does not know what data is important and should be annotated, one cannot collect and annotate it properly.

Although \textbf{exploratory research} may be possible and successful at uncovering correlations, evidence from a large birth cohort study\footnote{For example, the Avon Longitudinal Study of Parents and Children (ALSPAC), also known as Children of the 90s, which is a world-leading birth cohort study, charting the health of 14,500 families in the Bristol area (\url{http://www.bristol.ac.uk/alspac/}).} suggests that the approach ultimately leads to unsatisfactory outcomes.
It also has to be noted that for \textbf{smaller projects and time scales} (such as a PhD), a clear research direction is absolutely needed.

\subsubsection{Miscellaneous}
Apart from the above results, the following problems and ideas were discussed: 
\begin{itemize}
\item \textbf{two annotation groups:} a possible annotation strategy for live and fast paced activities could be that one group of annotators is used for detecting start / end of actions, and one other group for assigning the type of activity;
\item \textbf{making annotation attractive:} annotation with the current techniques and tools can often be a painful process. For example, researchers often experience that ``student annotators do not want to have anything to do with us anymore after the annotation'', which is especially problematic when you have put potentially considerable time and effort into training them. This rises the open question of how to motivate them to continue working on the annotation or how to make this process more attractive for them. One solution, tested by the University of Rostock, is to employ non-computer science students, preferably students with psychology or social sciences background. Such students already have experience in transcribing the data from large studies and are willing to annotate the data as opposed to computer science students who are more interested in implementation of the system that uses the annotation. What is more, computer science students could potentially be biased as they already have understanding of how the system should later work and be implemented, so they might make decisions based on the system's functionality / realisation.
%
%
\item \textbf{selecting appropriate sensors:} the question of whether external accelerometers are the best choice for movement data collection was discussed. Study designers should consider alternatives such as smart phones or smart watches that can be and are usually worn without constraints and that are part of the everyday life;
\item \textbf{is annotation needed:} changes in behaviours could potentially be detected based on the change in sensor data, which does not require ground-truth annotations. Here once again the question of study purpose arises: do the study designers want to detect changes in behaviour or more complex behavioural constructs.
\end{itemize}


%
%

\section{Conclusion and Future Perspectives}
\label{sec:conclusion}

In this technical report we presented the results from the 1st ARDUOUS Workshop. 
We attempted to identify relevant challenges in annotation of user data as well as potential solutions to these challenges. 
To that end we performed a live annotation session where we evaluated the annotation of seven participants. 
We used the annotation session as a basis for a discussion session where we identified important challenges and discussed possible solutions to specific problems. 

In that sense, this report provides a roadmap (or roadmaps) to problems identified in today's annotation practices. 
We hope that this roadmap will help in improving future annotation tools and methodologies and thus improve the quality of ubiquitous systems. 
It has to be noted that the workshop addressed just a small fraction of the problems associated with labelling. 
We hope to continue this effort in the future ARDUOUS workshops by discussing and identifying more relevant challenges. 
In the next ARDUOUS Workshop we hope to address the problem of multi-user annotation, keeping tract of user identities, and tool automation. 

\section*{Acknowledgements}
We would like to acknowledge the precious contribution of Alison Burrows, Niall Twomey, Pete Woznowski, and Tom Diethe from the University of Bristol, Rachel King from the University of Reading, and Jesse Hoey from the University of Waterloo for helping us shape the workshop. 
We are also very grateful to all workshop participants for their contributions during the live annotation and discussion sessions and for providing us with invaluable insights about the challenges, potential risks, and possible solutions for producing high-quality annotation.

\bibliographystyle{abbrv}
\bibliography{ref}

\newpage
\begin{appendices}
\section{Annotation of User Data: Questionnaire}

		\subsection{General questions}
			
			\begin{description}
			\item \textit{What is your age?} (in years)
			\item  \framebox{} under 20 \hspace{0.3cm}  \framebox{} 20 -- 30 \hspace{0.3cm} \framebox{} 30 -- 40 \hspace{0.3cm} \framebox{} 40 -- 50 \hspace{0.3cm} \framebox{} 50 -- 60 \hspace{0.3cm} \framebox{} above 60  
			\item
			\item \textit{What is your gender?}
			\item \framebox{} male  
			\item \framebox{} female
			\item \framebox{} other 
			\item
			\item \textit{Have you ever annotated user data?}
			\item \framebox{} yes, I often annotate user data  
			\item \framebox{} yes, I annotated several datasets
			\item \framebox{} yes, I annotated a large dataset once 
			\item \framebox{} yes, I annotated a small dataset once
			\item \framebox{} no, I have never annotated user data before
			\item 
			\item \textit{In which field of computer science do you feel you have solid knowledge?}
			\item ..........................................................................................................................
			\item
			\item \textit{If relevant, with what kind of annotation tools have you worked before?}			\item..........................................................................................................................
			\item ..........................................................................................................................
			\item ..........................................................................................................................
			\item
			\end{description}

\subsection{Live Annotation}

	\subsubsection{Single-user behaviour with simple action schema}

			\begin{description}
			\item \textit{How many problems occurred during this annotation?}
			\item ..........................................................................................................................
			\item ..........................................................................................................................
			\item 
			\item \textit{How many problems did you resolve during this annotation?}
			\item ..........................................................................................................................
			\item ..........................................................................................................................
			\item 
			\item \textit{Was it clear to you what was the meaining of the annotation labels?}
			\item \framebox{} very clear  \hspace{0.3cm} \framebox{} clear \hspace{0.3cm} \framebox{} relatively clear \hspace{0.3cm} \framebox{} unclear \hspace{0.3cm} \framebox{} very unclear
			\item
			\item \textit{Was it easy for you to recognise the annotation labels in the video?}
			\item \framebox{} very easy  \hspace{0.3cm} \framebox{} easy \hspace{0.3cm} \framebox{} relatively easy \hspace{0.3cm} \framebox{} difficult \hspace{0.3cm} \framebox{} very dificult
			\item 
			\item \textit{Did the annotation schema cover all observed actions?}
			\item \framebox{} yes, it covered all actions  \hspace{0.3cm} \framebox{} yes, it covered most actions \hspace{0.3cm} \framebox{} it covered some of the actions \hspace{0.3cm} \framebox{} no, most of the actions were missing \hspace{0.3cm} \framebox{} no, all of the actions were missing
			\item  
			\end{description}

	\subsubsection{Single-user behaviour with complex action schema}		
			\begin{description}
			\item \textit{How many problems occurred during this annotation?}
			\item ..........................................................................................................................
			\item ..........................................................................................................................
			\item 
			\item \textit{How many problems did you resolve during this annotation?}
			\item ..........................................................................................................................
			\item ..........................................................................................................................
			\item 
			\item \textit{Was it clear to you what was the meaining of the annotation labels?}
			\item \framebox{} very clear  \hspace{0.3cm} \framebox{} clear \hspace{0.3cm} \framebox{} relatively clear \hspace{0.3cm} \framebox{} unclear \hspace{0.3cm} \framebox{} very unclear
			\item 
			\item \textit{Was it easy for you to recognise the annotation labels in the video?}
			\item \framebox{} very clear  \hspace{0.3cm} \framebox{} clear \hspace{0.3cm} \framebox{} relatively clear \hspace{0.3cm} \framebox{} unclear \hspace{0.3cm} \framebox{} very unclear
			\item 
			\item \textit{Did the annotation schema cover all observed actions?}
			\item \framebox{} yes, it covered all actions  \hspace{0.3cm} \framebox{} yes, it covered most actions \hspace{0.3cm} \framebox{} it covered some of the actions \hspace{0.3cm} \framebox{} no, most of the actions were missing \hspace{0.3cm} \framebox{} no, all of the actions were missing
			\item  
			\end{description}
			
	\subsubsection{Multi-user behaviour without parallel actions}		
			
			\begin{description}
			\item \textit{How many problems occurred during this annotation?}
			\item ..........................................................................................................................
			\item ..........................................................................................................................
			\item 
			\item \textit{How many problems did you resolve during this annotation?}
			\item ..........................................................................................................................
			\item ..........................................................................................................................
			\item 
			\item \textit{Was it clear to you what was the meaining of the annotation labels?}
			\item \framebox{} very clear  \hspace{0.3cm} \framebox{} clear \hspace{0.3cm} \framebox{} relatively clear \hspace{0.3cm} \framebox{} unclear \hspace{0.3cm} \framebox{} very unclear
			\item 
			\item \textit{Was it easy for you to recognise the annotation labels in the video?}
			\item \framebox{} very clear  \hspace{0.3cm} \framebox{} clear \hspace{0.3cm} \framebox{} relatively clear \hspace{0.3cm} \framebox{} unclear \hspace{0.3cm} \framebox{} very unclear
			\item 
			\item \textit{Did the annotation schema cover all observed actions?}
			\item \framebox{} yes, it covered all actions  \hspace{0.3cm} \framebox{} yes, it covered most actions \hspace{0.3cm} \framebox{} it covered some of the actions \hspace{0.3cm} \framebox{} no, most of the actions were missing \hspace{0.3cm} \framebox{} no, all of the actions were missing
			\item  
			\end{description}
			
	\subsubsection{Multi-user behaviour with parallel actions}	
			\begin{description}
			\item \textit{How many problems occurred during this annotation?}
			\item ..........................................................................................................................
			\item ..........................................................................................................................
			\item 
			\item \textit{How many problems did you resolve during this annotation?}
			\item ..........................................................................................................................
			\item ..........................................................................................................................
			\item 
			\item \textit{Was it clear to you what was the meaining of the annotation labels?}
			\item \framebox{} very clear  \hspace{0.3cm} \framebox{} clear \hspace{0.3cm} \framebox{} relatively clear \hspace{0.3cm} \framebox{} unclear \hspace{0.3cm} \framebox{} very unclear
			\item
			\item \textit{Was it easy for you to recognise the annotation labels in the video?}
			\item \framebox{} very clear  \hspace{0.3cm} \framebox{} clear \hspace{0.3cm} \framebox{} relatively clear \hspace{0.3cm} \framebox{} unclear \hspace{0.3cm} \framebox{} very unclear
			\item 
			\item \textit{Did the annotation schema cover all observed actions?}
			\item \framebox{} yes, it covered all actions  \hspace{0.3cm} \framebox{} yes, it covered most actions \hspace{0.3cm} \framebox{} it covered some of the actions \hspace{0.3cm} \framebox{} no, most of the actions were missing \hspace{0.3cm} \framebox{} no, all of the actions were missing
			\item  
			\end{description}
			
			
\subsection{Annotation Tool}	

	\subsubsection{MMIS annotation tool}	
	
\begin{description}
	\item \textit{It was simple to use the tool.}
			\item \framebox{} strongly disagree  \hspace{0.3cm} \framebox{} disagree \hspace{0.3cm} \framebox{} moderately disagree \hspace{0.3cm} \framebox{} neutral \hspace{0.3cm} \framebox{} moderately agree  \hspace{0.3cm} \framebox{} agree  \hspace{0.3cm} \framebox{} strongly agree
			\item 
			\item \textit{I can effectively annotate using the tool.}
			\item \framebox{} strongly disagree  \hspace{0.3cm} \framebox{} disagree \hspace{0.3cm} \framebox{} moderately disagree \hspace{0.3cm} \framebox{} neutral \hspace{0.3cm} \framebox{} moderately agree  \hspace{0.3cm} \framebox{} agree  \hspace{0.3cm} \framebox{} strongly agree
			\item 
			\item \textit{I am able to complete the annotation quickly using the tool.}
			\item \framebox{} strongly disagree  \hspace{0.3cm} \framebox{} disagree \hspace{0.3cm} \framebox{} moderately disagree \hspace{0.3cm} \framebox{} neutral \hspace{0.3cm} \framebox{} moderately agree  \hspace{0.3cm} \framebox{} agree  \hspace{0.3cm} \framebox{} strongly agree
			\item 
			\item \textit{It was easy to learn to use the tool.}
			\item \framebox{} strongly disagree  \hspace{0.3cm} \framebox{} disagree \hspace{0.3cm} \framebox{} moderately disagree \hspace{0.3cm} \framebox{} neutral \hspace{0.3cm} \framebox{} moderately agree  \hspace{0.3cm} \framebox{} agree  \hspace{0.3cm} \framebox{} strongly agree
			\item 
			\item \textit{The tool gives error messages that clearly tell me how to fix problems.}
			\item \framebox{} strongly disagree  \hspace{0.3cm} \framebox{} disagree \hspace{0.3cm} \framebox{} moderately disagree \hspace{0.3cm} \framebox{} neutral \hspace{0.3cm} \framebox{} moderately agree  \hspace{0.3cm} \framebox{} agree  \hspace{0.3cm} \framebox{} strongly agree
			\item 	
			\item \textit{Whenever I make a mistake using the tool, I recover easily and quickly.}
			\item \framebox{} strongly disagree  \hspace{0.3cm} \framebox{} disagree \hspace{0.3cm} \framebox{} moderately disagree \hspace{0.3cm} \framebox{} neutral \hspace{0.3cm} \framebox{} moderately agree  \hspace{0.3cm} \framebox{} agree  \hspace{0.3cm} \framebox{} strongly agree
			\item 	
			\item \textit{The organisation of information on the screen is clear.}
			\item \framebox{} strongly disagree  \hspace{0.3cm} \framebox{} disagree \hspace{0.3cm} \framebox{} moderately disagree \hspace{0.3cm} \framebox{} neutral \hspace{0.3cm} \framebox{} moderately agree  \hspace{0.3cm} \framebox{} agree  \hspace{0.3cm} \framebox{} strongly agree
			\item 
			\item \textit{The interface of the tool is pleasant.}
			\item \framebox{} strongly disagree  \hspace{0.3cm} \framebox{} disagree \hspace{0.3cm} \framebox{} moderately disagree \hspace{0.3cm} \framebox{} neutral \hspace{0.3cm} \framebox{} moderately agree  \hspace{0.3cm} \framebox{} agree  \hspace{0.3cm} \framebox{} strongly agree
			\item 
			\item \textit{The tool is suitable for online annotation.}
			\item \framebox{} strongly disagree  \hspace{0.3cm} \framebox{} disagree \hspace{0.3cm} \framebox{} moderately disagree \hspace{0.3cm} \framebox{} neutral \hspace{0.3cm} \framebox{} moderately agree  \hspace{0.3cm} \framebox{} agree  \hspace{0.3cm} \framebox{} strongly agree
			\item 
			\item \textit{The tool is suitable for offline annotation.}
			\item \framebox{} strongly disagree  \hspace{0.3cm} \framebox{} disagree \hspace{0.3cm} \framebox{} moderately disagree \hspace{0.3cm} \framebox{} neutral \hspace{0.3cm} \framebox{} moderately agree  \hspace{0.3cm} \framebox{} agree  \hspace{0.3cm} \framebox{} strongly agree
			\item 
			\item \textit{This tool has all the functions and capabilities I expect it to have.}
			\item \framebox{} strongly disagree  \hspace{0.3cm} \framebox{} disagree \hspace{0.3cm} \framebox{} moderately disagree \hspace{0.3cm} \framebox{} neutral \hspace{0.3cm} \framebox{} moderately agree  \hspace{0.3cm} \framebox{} agree  \hspace{0.3cm} \framebox{} strongly agree
			\item 
			\item \textit{List the functions and capabilities that are missing (if applicable).}
			\item ..........................................................................................................................
			\item ..........................................................................................................................
			\item ..........................................................................................................................
			\item ..........................................................................................................................
			\item 
			\item \textit{Overall, I am satisfied with the tool.}
			\item \framebox{} strongly disagree  \hspace{0.3cm} \framebox{} disagree \hspace{0.3cm} \framebox{} moderately disagree \hspace{0.3cm} \framebox{} neutral \hspace{0.3cm} \framebox{} moderately agree  \hspace{0.3cm} \framebox{} agree  \hspace{0.3cm} \framebox{} strongly agree
			\item 
			\item \textit{List the most negative aspects of the system.}
			\item ..........................................................................................................................
			\item ..........................................................................................................................
			\item ..........................................................................................................................
			\item ..........................................................................................................................
			\item 
			\item \textit{List the most positive aspects of the system.}
			\item ..........................................................................................................................
			\item ..........................................................................................................................
			\item ..........................................................................................................................
			\item ..........................................................................................................................
			\item 	
	\end{description}

	\subsubsection{SPHERE annotation tool}

	\begin{description}
	\item \textit{It was simple to use the tool.}
			\item \framebox{} strongly disagree  \hspace{0.3cm} \framebox{} disagree \hspace{0.3cm} \framebox{} moderately disagree \hspace{0.3cm} \framebox{} neutral \hspace{0.3cm} \framebox{} moderately agree  \hspace{0.3cm} \framebox{} agree  \hspace{0.3cm} \framebox{} strongly agree
			\item 
			\item \textit{I can effectively annotate using the tool.}
			\item \framebox{} strongly disagree  \hspace{0.3cm} \framebox{} disagree \hspace{0.3cm} \framebox{} moderately disagree \hspace{0.3cm} \framebox{} neutral \hspace{0.3cm} \framebox{} moderately agree  \hspace{0.3cm} \framebox{} agree  \hspace{0.3cm} \framebox{} strongly agree
			\item 
			\item \textit{I am able to complete the annotation quickly using the tool.}
			\item \framebox{} strongly disagree  \hspace{0.3cm} \framebox{} disagree \hspace{0.3cm} \framebox{} moderately disagree \hspace{0.3cm} \framebox{} neutral \hspace{0.3cm} \framebox{} moderately agree  \hspace{0.3cm} \framebox{} agree  \hspace{0.3cm} \framebox{} strongly agree
			\item 
			\item \textit{It was easy to learn to use the tool.}
			\item \framebox{} strongly disagree  \hspace{0.3cm} \framebox{} disagree \hspace{0.3cm} \framebox{} moderately disagree \hspace{0.3cm} \framebox{} neutral \hspace{0.3cm} \framebox{} moderately agree  \hspace{0.3cm} \framebox{} agree  \hspace{0.3cm} \framebox{} strongly agree
			\item 
			\item \textit{The tool gives error messages that clearly tell me how to fix problems.}
			\item \framebox{} strongly disagree  \hspace{0.3cm} \framebox{} disagree \hspace{0.3cm} \framebox{} moderately disagree \hspace{0.3cm} \framebox{} neutral \hspace{0.3cm} \framebox{} moderately agree  \hspace{0.3cm} \framebox{} agree  \hspace{0.3cm} \framebox{} strongly agree
			\item 	
			\item \textit{Whenever I make a mistake using the tool, I recover easily and quickly.}
			\item \framebox{} strongly disagree  \hspace{0.3cm} \framebox{} disagree \hspace{0.3cm} \framebox{} moderately disagree \hspace{0.3cm} \framebox{} neutral \hspace{0.3cm} \framebox{} moderately agree  \hspace{0.3cm} \framebox{} agree  \hspace{0.3cm} \framebox{} strongly agree
			\item 	
			\item \textit{The organisation of information on the screen is clear.}
			\item \framebox{} strongly disagree  \hspace{0.3cm} \framebox{} disagree \hspace{0.3cm} \framebox{} moderately disagree \hspace{0.3cm} \framebox{} neutral \hspace{0.3cm} \framebox{} moderately agree  \hspace{0.3cm} \framebox{} agree  \hspace{0.3cm} \framebox{} strongly agree
			\item 
			\item \textit{The interface of the tool is pleasant.}
			\item \framebox{} strongly disagree  \hspace{0.3cm} \framebox{} disagree \hspace{0.3cm} \framebox{} moderately disagree \hspace{0.3cm} \framebox{} neutral \hspace{0.3cm} \framebox{} moderately agree  \hspace{0.3cm} \framebox{} agree  \hspace{0.3cm} \framebox{} strongly agree
			\item 
			\item \textit{The tool is suitable for online annotation.}
			\item \framebox{} strongly disagree  \hspace{0.3cm} \framebox{} disagree \hspace{0.3cm} \framebox{} moderately disagree \hspace{0.3cm} \framebox{} neutral \hspace{0.3cm} \framebox{} moderately agree  \hspace{0.3cm} \framebox{} agree  \hspace{0.3cm} \framebox{} strongly agree
			\item 
			\item \textit{The tool is suitable for offline annotation.}
			\item \framebox{} strongly disagree  \hspace{0.3cm} \framebox{} disagree \hspace{0.3cm} \framebox{} moderately disagree \hspace{0.3cm} \framebox{} neutral \hspace{0.3cm} \framebox{} moderately agree  \hspace{0.3cm} \framebox{} agree  \hspace{0.3cm} \framebox{} strongly agree
			\item 
			\item \textit{This tool has all the functions and capabilities I expect it to have.}
			\item \framebox{} strongly disagree  \hspace{0.3cm} \framebox{} disagree \hspace{0.3cm} \framebox{} moderately disagree \hspace{0.3cm} \framebox{} neutral \hspace{0.3cm} \framebox{} moderately agree  \hspace{0.3cm} \framebox{} agree  \hspace{0.3cm} \framebox{} strongly agree
			\item 
			\item \textit{List the functions and capabilities that are missing (if applicable).}
			\item ..........................................................................................................................
			\item ..........................................................................................................................
			\item ..........................................................................................................................
			\item ..........................................................................................................................
			\item 
			\item \textit{Overall, I am satisfied with the tool.}
			\item \framebox{} strongly disagree  \hspace{0.3cm} \framebox{} disagree \hspace{0.3cm} \framebox{} moderately disagree \hspace{0.3cm} \framebox{} neutral \hspace{0.3cm} \framebox{} moderately agree  \hspace{0.3cm} \framebox{} agree  \hspace{0.3cm} \framebox{} strongly agree
			\item 
			\item \textit{List the most negative aspects of the system.}
			\item ..........................................................................................................................
			\item ..........................................................................................................................
			\item ..........................................................................................................................
			\item ..........................................................................................................................
			\item 
			\item \textit{List the most positive aspects of the system.}
			\item ..........................................................................................................................
			\item ..........................................................................................................................
			\item ..........................................................................................................................
			\item ..........................................................................................................................
	
	\end{description}		
\end{appendices}

\end{document}